# Cascade of correlated electron states in a kagome superconductor CsV$_3$Sb$_5$

He Zhao[1], Hong Li[1], Brenden R. Ortiz[2], Samuel M. L. Teicher[2], Taka Park[3], Mengxing Ye[3], Ziqiang Wang[1], Leon Balents[3], Stephen D. Wilson[2] and Ilija Zeljkovic[1,*]

[1] *Department of Physics, Boston College, Chestnut Hill, MA 02467, USA*

[2] *Materials Department and California Nanosystems Institute, University of California Santa Barbara, Santa Barbara, California 93106, USA*

[3] *Kavli Institute for Theoretical Physics, University of California, Santa Barbara, Santa Barbara, California 93106, USA*

*Correspondence: ilija.zeljkovic@bc.edu

**The kagome lattice of transition metal atoms provides an exciting platform to study electronic correlations in the presence of geometric frustration and nontrivial band topology, which continues to bear surprises. In this work, using spectroscopic imaging scanning tunneling microscopy, we discover a cascade of different symmetry-broken electronic states as a function of temperature in a new kagome superconductor, CsV$_3$Sb$_5$. At a temperature far above the superconducting transition $T_c$ ~ 2.5 K, we reveal a tri-directional charge order with a $2a_0$ period that breaks the translation symmetry of the lattice. As the system is cooled down towards $T_c$, we observe a prominent V-shape spectral gap opening at the Fermi level and an additional breaking of the six-fold rotation symmetry, which persists through the superconducting transition. This rotation symmetry breaking is observed as the emergence of an additional $4a_0$ unidirectional charge order and strongly anisotropic scattering in differential conductance maps. The latter can be directly attributed to the orbital-selective renormalization of the V kagome bands. Our experiments reveal a complex landscape of electronic states that can co-exist on a kagome lattice, and provide intriguing parallels to high-$T_c$ superconductors and twisted bilayer graphene.**

Quantum solids composed of atoms arranged on a lattice of corner-sharing triangles (kagome lattice) are a fascinating playground for the exploration of novel correlated and topological electronic phenomena [1–4]. Due to their intrinsic geometric frustration, kagome systems are predicted to host to a slew of exotic electronic states [5–18], such as bond and charge ordering [7,8,10,16–18], spin liquid phases [5,15] and chiral superconductivity [9,10,17]. The majority of the experimental efforts thus far have focused on transition-metal kagome magnets, for example Co$_3$Sn$_2$S$_2$ [19–23], FeSn [24,25] and Fe$_3$Sn$_2$ [26,27], in which different forms of magnetism dominate the low-temperature electronic ground state. Electronic correlations in the absence of magnetic ordering could in principle favor the emergence of new symmetry-broken electronic states, but this has been difficult to explore in many of the existing kagome materials due to a tendency towards magnetic ordering.

$A$V$_3$Sb$_5$ ($A$=K, Rb, Cs) is a recently discovered class of kagome metals that does not exhibit resolvable magnetic order [28,29]. This family of materials has already shown a glimpse of unusual electronic behavior in a topologically nontrivial setting, such as the large anomalous Hall response born from substantial Berry curvature [30] and a rare occurrence of superconductivity in a kagome system [29,31]. Theoretical band structure calculations of $A$V$_3$Sb$_5$ indicate that the band structure is characterized by a non-trivial topological invariant [29,31], which combined with emergent superconductivity, draws an interesting parallel to the intensely studied topological metals in the family of Fe-based high-$T_c$ superconductors [32,33]. The large density of states due to a van Hove singularity that is located in the vicinity of the Fermi level, and quasi-1D regions of the Fermi surface [29] also provide an ideal playground for the search of elusive correlated states on a kagome lattice. While theory predicts numerous possibilities for translation and rotation symmetry breaking of the kagome lattice electronic structure [10,17,18], their experimental realization has been challenging. In this work, using spectroscopic-imaging scanning tunneling microcopy (SI-STM), we discover a cascade of symmetry-broken phases in a kagome superconductor CsV$_3$Sb$_5$ as a function of temperature, detectable as charge ordering at different wave vectors and an anisotropic quasiparticle scattering signature. These phases develop in the normal state, first breaking the translation and subsequently also the rotation symmetry of the lattice, and persist below the superconducting $T_c$. Our experiments demonstrate that superconductivity in CsV$_3$Sb$_5$ emerges from, and coexists with, an electronic state with an intrinsically broken rotational symmetry. This could have strong implications on the superconducting order parameter in this family of kagome superconductors, where transport experiments recently suggested the possibility of unconventional pairing [34,35].

CsV$_3$Sb$_5$ is a layered superconductor ($T_c$~2.5 K) [29], with a hexagonal crystal structure ($a$=$b$=5.4 Å, $c$=9 Å) composed of V-Sb sheets stacked between complete Cs layers [28,29] (Fig. 1a,b). Each V-Sb sheet consists of a kagome network of V atoms, interlaced with a hexagonal lattice of Sb. Due to a strong chemical bond expected between V and Sb atoms, the material is anticipated to cleave between the V-Sb sheet and the Cs layer. Consistent with this expectation, STM topographs show two main types of atomically-ordered surface morphologies: a hexagonal lattice that we attribute to the Cs layer, and a honeycomb-like surface structure, which we ascribe to the Sb layer (Fig. 1e,f). We find that a complete Cs surface layer is unstable, often leading to randomly distributed Cs atoms prone to clustering (Fig. 1(c,d), Fig. S1). The Sb-terminated layer is on the other hand robust, and we can routinely locate large pristine areas for imaging (Fig. S2).

In the remainder of the work, we focus on the Sb-terminated surface, located directly above the V kagome layer. Differential conductance d$I$/d$V$ spectra, proportional to the local density of states, exhibit several distinct features, which are denoted by arrows in Fig. 2b. By comparison to theoretical band structure calculations, which show a good agreement with angle-resolved photoemission spectroscopy (ARPES) [29], we can identify the two spectral peaks in d$I$/d$V$ spectra below the Fermi level as van Hove singularities at the M point. The local minimum between them is likely associated with the Dirac point at K (Fig. 2a,b). To gain further insight into the electronic structure of our system, we use quasiparticle interference (QPI) imaging, rooted in the elastic scattering and interference of electrons visible as static modulations in d$I$/d$V$(**r**,$V$) maps. Aside from atomic Bragg peaks $\mathbf{Q}_{Bragg}^i$ ($i = a, b, c$), the Fourier transforms (FTs) of d$I$/d$V$(**r**,$V$) maps show an isotropic scattering vector $q_1$ near the FT center in momentum-transfer space ($q$-space) (Fig. 2c, Fig. S3). This vector is present in our data across a wide range of energies crossing the Fermi level. By comparing its dispersion as a function of energy to theoretical band structure, we identify $q_1$ as the scattering vector connecting different states on the constant energy contour of the pocket centered at Γ (Fig. 2a,c,d). The Fermi vector $k_f \sim 0.18$ Å$^{-1}$ obtained from our data ($q_1(E = 0) = 2k_f$) is nearly identical to that measured by ARPES [29]. This agreement, together with the identification of spectral features in d$I$/d$V$ spectra (Fig. 2a,b), demonstrates the consistency between the electronic band structures measured by ARPES, STM and theory.

Average d$I$/d$V$ spectra also show a partial suppression of the spectral weight in the vicinity of the Fermi level (inset in Fig. 2b). This is suggestive of the existence of an ordered electronic state, possibly related to the transport anomaly at $T^* = 94$ K attributed to a CDW like instability [29]. We investigate this by acquiring temperature-dependent STM measurements, starting at a high temperature below $T^*$. We find that STM topographs at ~60 K exhibit a 2 x 2 superstructure, which breaks the translational symmetry of the lattice (Fig. 3a). In Fourier space, this pattern gives rise to a set of wave vectors at exactly $q_{3Q-CO}^i = \frac{1}{2} Q_{Bragg}^i$ ($i = a, b, c$) (Fig. 2c, Fig. 3d-g). The vectors are extremely localized in reciprocal space (~1-2 pixels or ~ 0.006 - 0.012 Å$^{-1}$) and their positions do not disperse with energy (Fig. 2c, Fig. 3g, Fig. S3). Based on this, we attribute these features to a static charge order with a $2a_0$ period, propagating along all 3 lattice directions (3Q-CO). This interpretation is bolstered by the X-ray diffraction measurements that detected peaks at the same wave vectors emerging below $T^*$ [29] and a recent report of charge ordering at the same wave vector in the cousin compound KV$_3$Sb$_5$ [36].

As the system is cooled down further, the electronic structure of CsV$_3$Sb$_5$ begins to display a pronounced unidirectional character. While 3Q-CO charge order is still present at ~4.5 K, STM topographs exhibit another periodic modulation with a $4a_0$ wave length propagating along only one lattice direction (Fig. 3b,c). This charge "stripe" modulation corresponds to the FT wave vector $q_{1Q-CO} = 0.23\, Q_{Bragg}$ (Fig. 3d,f,g). The small deviation between this average vector magnitude $q_{1Q-CO} = 0.23\, Q_{Bragg}$ and the commensurate location 0.25 $Q_{Bragg}$ can be attributed to phase slips (inset in Fig. 3c). The wave vector $q_{1Q-CO}$ is again non-dispersive (Fig. 3g), and thus consistent with a unidirectional charge order (1Q-CO), which has not been experimentally observed in any kagome system to-date. Upon a closer inspection of FT intensities along the same direction, we notice several faint peaks, which are equally spaced by ~0.05 |$Q_{Bragg}$| around dominant wave vectors (Fig. S4). These cannot be explained by a linear superposition of $Q_{Bragg}$, $q_{1Q-CO}$ and $q_{3Q-CO}^i$, and may be indicative of another charge ordering state with the wave vector ~0.05 |$Q_{Bragg}$| coupling to other peaks via satellite reflections.

In addition to static charge ordering, our SI-STM measurements reveal another intriguing aspect of the electronic structure. At low energies, FTs of d$I$/d$V$(**r**, $V$) maps display long parallel features in reciprocal space, which we label as $q_2$ and $q'_2$ based on their $q$-space position (Fig. 4b, Fig. S5). Each $q_2$ and $q'_2$ set of wave vectors consists of multiple parallel stripes in $q$-space. Importantly, the separation between these $q$-space stripes disperses noticeably over a narrow energy range where the feature is detected (Fig. 4c). This observation demonstrates that $q_2$ and $q'_2$ are related to elastic scattering between different points on the constant-energy contour (CEC) – the separation between the **q**-space stripe features at different energies will evolve concomitantly with the change in the CEC in **k**-space. To understand the origin of scattering in more detail, we examine the morphology of the Fermi surface. The CEC at zero energy consists of parallel quasi-1D bands along $M_1$-$M_2$ direction originating from V orbitals (inset in Fig. 4a, Supplementary Information 1). Scattering between them could naturally give rise to the unidirectional QPI signature observed. To visualize this, we simulate the QPI signature based on the CEC at zero energy approximately expected from ARPES, where **q₂** and **q′₂** can be beautifully reproduced (Fig. 4a, Supplementary Information 2).

Remarkably, **q₂** and **q′₂** are only observed along a single lattice vector parallel to $\mathbf{q_{1Q-CO}}$, while they are notably absent at the equivalent **q**-space positions along the other two lattice directions (dashed rectangles in Fig. 4b). This provides further evidence that the electronic band structure of CsV$_3$Sb$_5$ breaks the six-fold rotational symmetry of the lattice. We rule out STM tip anisotropy affecting our observations, as the unidirectional electronic signature rotates across a domain boundary (Fig. 4f,g). As we previously discussed, the $C_2$-symmetric signal in the FTs of d$I$/d$V$(**r**, $V$) maps originates from scattering between the states characterized by V orbitals (Fig. 4a, Fig. S10). This observation suggests that although our measurements are performed on top of the Sb-terminated surface, the rotational symmetry breaking observed is intrinsic to the V kagome layer itself. It is also important to note that it is band-specific. The QPI vector **q₁** related to the Sb pocket at Γ does not show any noticeable $q$-space anisotropy within the same energy range (Fig. 4d).

By a closer inspection of **q₂** and **q′₂** in QPI maps away from the Fermi energy, we find that these vectors get suppressed above ~12 meV, around the same energy at which the equivalent vectors along the two other lattice directions emerge (Fig. 4d,e). This observation points towards a strong energy-dependent orbital renormalization, which is reminiscent of the orbital- and energy-dependent quasiparticle spectral weight in the electronic nematic state of Fe-based superconductors [37]. Rotation symmetry breaking could in principle be explained by either strong nematic susceptibility pinned by a small accidental strain during the sample cooldown, or an intrinsic symmetry-broken electronic phase in the absence of strain. In strong support of the latter, we point to the following: First, symmetry-breaking QPI has been consistently observed over > 100 nm square regions in multiple different crystals. Second, we can observe electronic domains in the absence of obvious structural imperfections (Fig. 4f,g, Fig. S6). Third, no rotational symmetry breaking occurs at ~60 K, the temperature at which the majority of the sample thermal contraction would have occurred, and yet a unidirectional modulation can be clearly seen at ~50 K (Fig. 3e). Any strain-induced effects over such narrow temperature range should be negligible. Lastly, repeated scanning at low junction resistance can occasionally modify the charge ordering direction near the domain boundaries, possibly due to the electric field of the tip (Fig. S6). For these reasons, our experiments present strong evidence for an intrinsic rotation symmetry breaking of the electronic structure of CsV$_3$Sb$_5$.

We establish a rich landscape of symmetry-broken states that can emerge and co-exist on a kagome lattice. We hypothesize that the 3Q-CO arises from the nesting between the saddle points at M, giving rise to a charge order with exactly 2$a_0$ wave length. The 4$a_0$ 1Q-CO on the other hand does not fit the existing theoretical models of charge ordering proposed to exist on a kagome lattice [13,14]. It poses an intriguing puzzle: canonically unidirectional CDWs are associated with a Peierls instability (at twice the Fermi wave vector) of quasi-1D bands; however, the wave vector of the 1Q-CO (along Γ-M) is perpendicular to the nesting vector connecting the quasi-1D V bands (along Γ-K). Given the existence of multiple bands crossing the Fermi level in this system, it is conceivable that the two phenomena – non-dispersive $4a_0$ 1Q-CO (Fig. 3c,g) and dispersive anisotropic quasiparticle scattering (Fig. 4b-e, Fig. S12) – are associated with different electronic bands. In this scenario, anisotropic QPI could be regarded as a $q$=0 state that only breaks the rotation symmetry of the lattice, and interpreted as a purely nematic order related to a V band not actively involved in the $4a_0$ order. Nematic ordering with a three-state Potts-nematic order parameter [38] has recently come into spotlight in a hexagonal Moire lattice of twisted bilayer graphene (TBG) [39–42], where electronic nematicity arises due to strong correlations within a partially-filled nearly flat band. In contrast to the nematicity in TBG however, rotation symmetry breaking of the hexagonal lattice in $CsV_3Sb_5$ clearly goes beyond a partial filling of localized bands, and it can be traced to delocalized electronic states of the V bands. Additional experiments and theoretical modeling will be necessary to establish the exact relationship between different electronic phases observed here.

Lastly, we note that sub-Kelvin SI-STM measurements reveal that symmetry-broken states persist well below the superconducting $T_c$ (Fig. S7), thus adding to the complexity of the $CsV_3Sb_5$ phase diagram. The co-existence of the $4a_0$ charge stripe order that breaks the rotational symmetry of the lattice, V-shape spectral gap opening in the normal state, and superconductivity presents a remarkable similarity to the observed electronic states in cuprate high-temperature superconductors [43]. It remains to be seen if the superconducting order parameter in $CsV_3Sb_5$ is also unconventional, as possibly indicated by recent transport measurements [34,35]. Future experiments should address the competition of different phases by a more detailed temperature-, energy- and doping-dependent measurements, while also searching for evidence of intrinsic topological superconductivity and Majorana modes expected to arise in this family of materials due to their intrinsic non-trivial band topology.

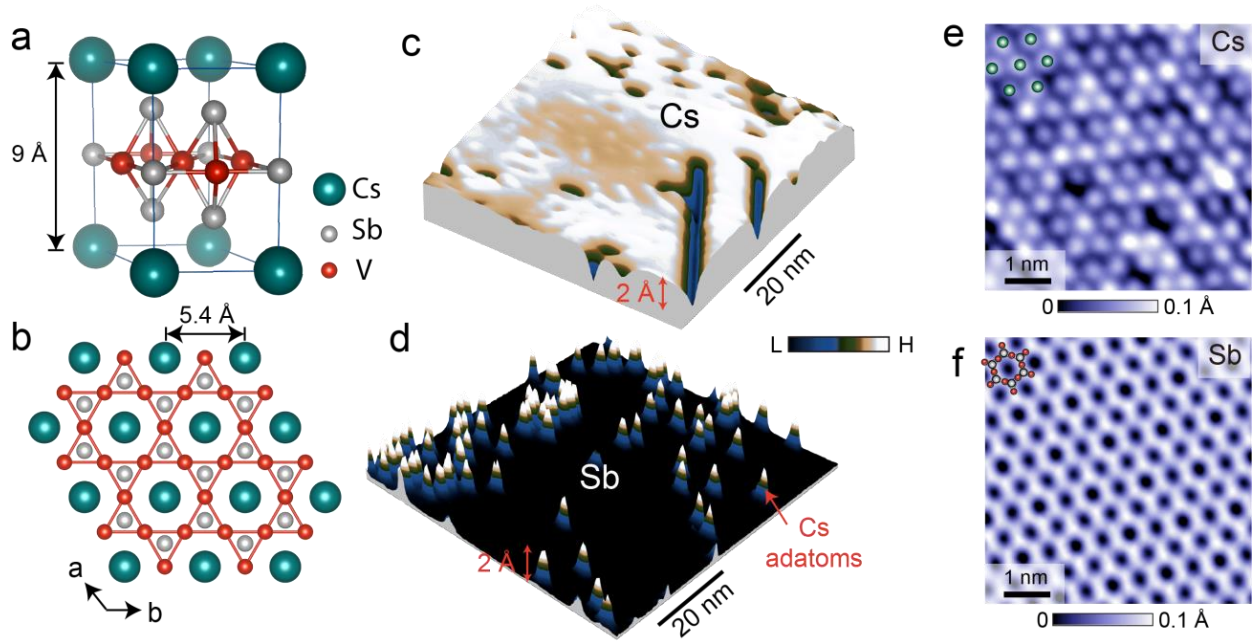

**Figure 1.** Surface identification. (a) 3D crystal structure of $CsV_3Sb_5$ viewed from the side. (b) Top-view of the crystal structure (*ab*-plane), showing the kagome lattice of the $V_3Sb$ layer and the triangular lattice of the Cs layer. (c,d) 3D portrayal of a large-scale morphology of (c) Cs and (d) Sb layers from STM topographs. (e,f) Atomically resolved scanning tunneling microscopy (STM) topographs of (e) Cs-terminated and (f) Sb-terminated surfaces, with atom ball models superimposed. Green, gray and red spheres in (a,b,e,f) denote Cs, Sb and V atoms, respectively. STM setup condition: (c) $V_{sample}$ = 200 mV, $I_{set}$ = 40 pA, T = 4.5 K; (d) $V_{sample}$ = 200 mV, $I_{set}$ = 10 pA, T = 4.5 K, (e) $V_{sample}$ = 20 mV, $I_{set}$ = 200 pA, T = 4.5 K, (f) $V_{sample}$ = 50 mV, $I_{set}$ = 30 pA, T = 59 K;

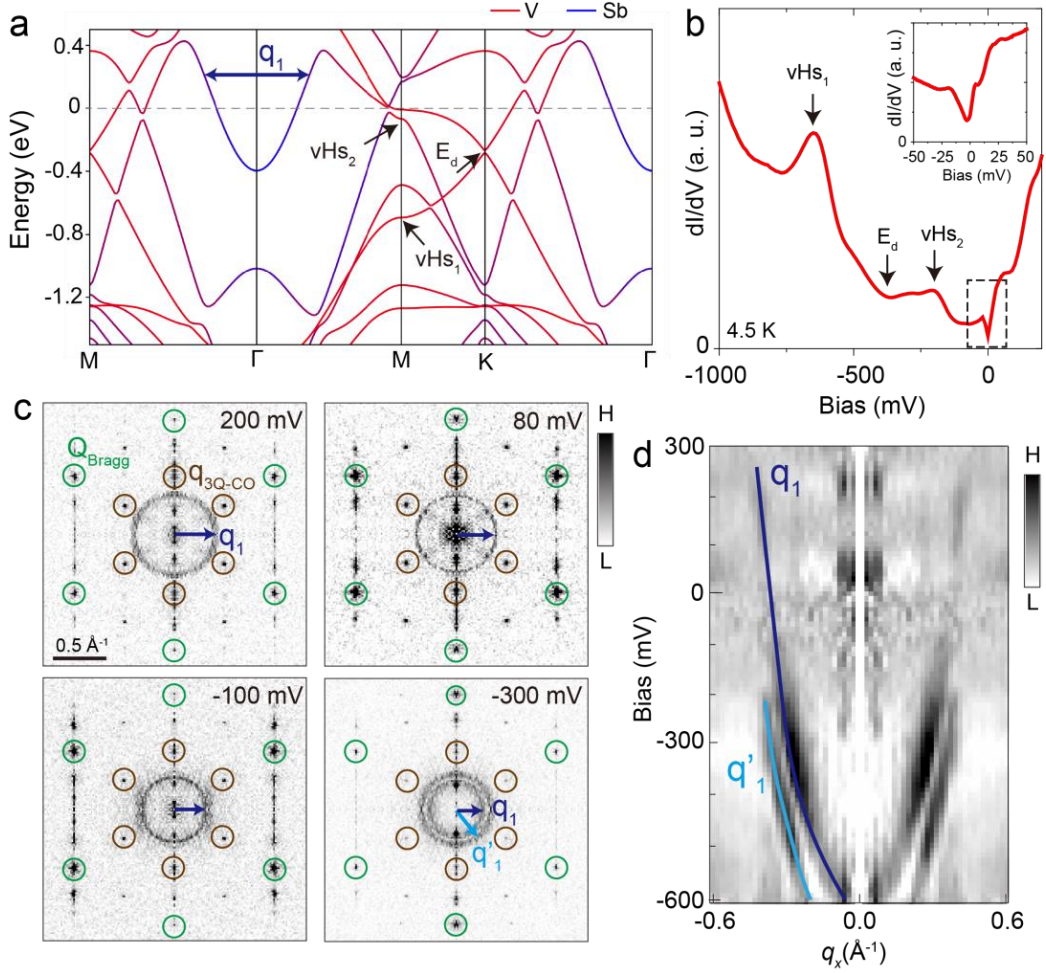

**Figure 2.** Large-scale electronic characterization. (a) Calculated band structure of $CsV_3Sb_5$, with two of the van Hove singularities ($vHs_{1,2}$) at M and one Dirac point ($E_d$) at K called out by arrows (Supplementary Information 1, Fig. S10). (b) Spatially averaged differential conductance ($dI/dV$) spectrum acquired over the Sb-terminated surface. The two peaks and the dip in the spectrum match the relative positions of the calculated $vHs_{1,2}$ and $E_d$, respectively, with only a small chemical potential shift. Inset in (b) displays the $dI/dV$ spectrum over a smaller energy range close to Fermi energy ($E_F$). (c) Representative two-fold symmetrized Fourier transforms (FTs) of $dI/dV(\mathbf{r},V)$ maps, taken over an identical region of the Sb surface, showing prominent circle-like scattering vector $\mathbf{q}_1$. The green circles indicate the atomic Bragg peaks, while the brown circles denote the charge order $\mathbf{q}_{3Q\text{-}CO}$. (d) Radially averaged linecut in two-fold symmetrized FTs of $L(\mathbf{r}, V)$ maps (defined as $dI/dV(\mathbf{r},V)/(I(\mathbf{r},V)/V)$). Dark (light) blue arrows and lines in (c,d) denote the most prominent dispersive scattering vectors $\mathbf{q}_1$ ($\mathbf{q}'_1$) in momentum-transfer space. While we cannot conclusively identify $\mathbf{q}'_1$, we hypothesize that it may be related to the same band as $\mathbf{q}_1$, and that due to a larger $\mathbf{k}_z$-dispersion at lower energies, the STM picks out dispersions related to two different, possibly extreme values of $\mathbf{k}_z$. To emphasize other features, the center pixel of FT linecut in (d) has been artificially suppressed. STM setup condition: (b) $V_{sample} = 200$ mV, $I_{set} = 500$pA, $V_{exc} = 5$ mV; inset in (b) $V_{sample} = 100$ mV, $I_{set} = 500$pA, $V_{exc} = 1$ mV; (c) $R_{tip\text{-}sample} = 0.5$ GOhms, $V_{exc} = 4$ mV; (d) $V_{sample} = 300$ mV, $I_{set} = 200$pA, $V_{exc} = 4$ mV. All data are taken at 4.5K.

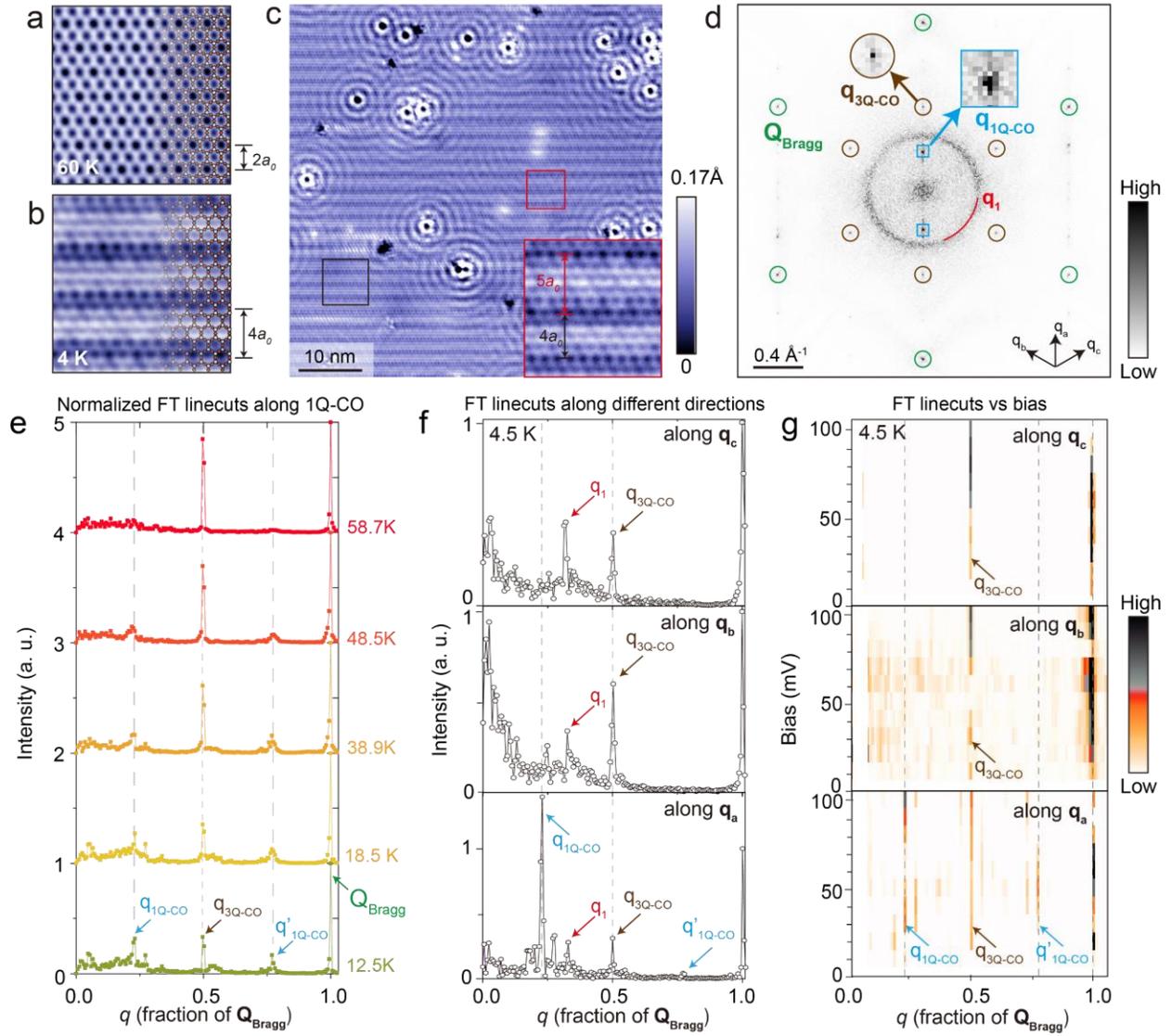

**Figure 3.** Charge ordering at low temperature. (a,b) Zoom-in comparison of atomically-resolved scanning tunneling microscopy (STM) topographs, taken over the Sb-terminated surface at (a) ~60 K and (b) ~4.5 K, showing different charge modulation patterns (see also Fig. S9). (c) Low-temperature STM topograph over a large region. Panel (b) is a close-up of the black squared region in the topograph in (c). Inset in (c) shows an example of a phase slip due to a $5a_0$ stripe charge modulation. The honeycomb Sb surface lattice structure is superimposed in (a,b). (d) The Fourier transform (FT) of STM topograph in (c). Green circles, brown circles and blue squares indicate the atomic Bragg peaks, 3Q-CO and 1Q-CO peaks, respectively. (e) Temperature dependent FT linecuts, along the 1Q-CO direction, all taken from data acquired over an identical region. Gray dash lines are visual guides showing a non-dispersive nature of the peaks. (f) FT linecuts along the three different lattice directions in (d). (g) FT linecuts of $L(\mathbf{r}, V)$ maps along three lattice directions at ~4.5 K as a function of bias. STM setup condition: (a) $V_{sample}$ = 50 mV, $I_{set}$ = 30 pA, T = 60 K; (b,c) $V_{sample}$ = -20 mV, $I_{set}$ = 20 pA, T = 4.5 K; (f) $V_{sample}$ = 100 mV, $I_{set}$ = 600 pA, $V_{exc}$ = 4 mV, T = 4.5 K; (g) $V_{sample}$ = -30 mV, $I_{set}$ = 20 pA.

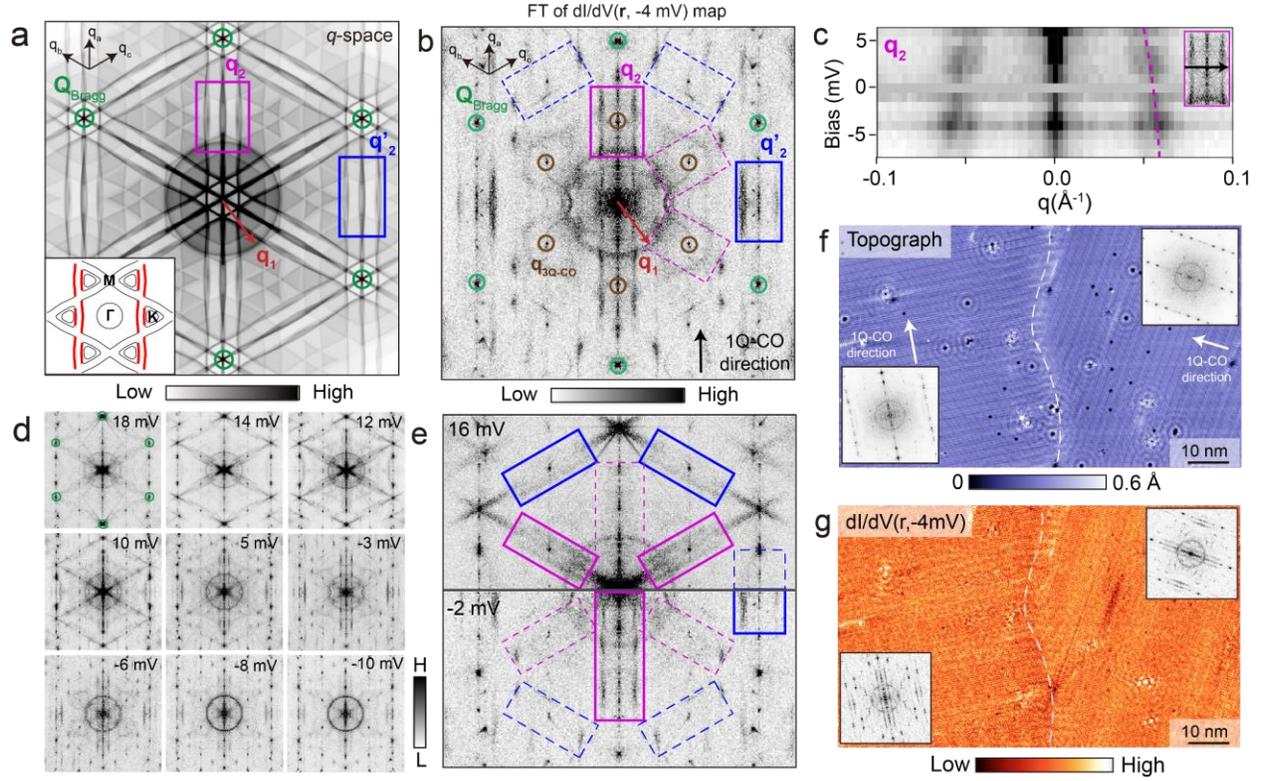

**Figure 4.** Visualizing rotational symmetry breaking in the quasiparticle interference (QPI) of $CsV_3Sb_5$. (a) Expected QPI pattern at zero energy, calculated by the autocorrelation of the schematic of the Fermi surface (inset in (a), Supplementary Information 2). Red highlighted parts in the inset in (a) denote the parts of the Fermi surface that lead to the anisotropic QPI patterns (Fig. S11). (b) Two-fold symmetrized Fourier transform (FT) of differential conductance d$I$/d$V$(**r**, -4 mV) map over the Sb-terminated surface. Stripe features propagating along the $q_a$ direction (solid rectangles labeled $q_2$ and $q'_2$), are absent in $q_b$ and $q_c$ lattice directions (dashed rectangles). (c) Linecuts of FTs of dI/dV(**r**, V) maps as a function of bias along the black arrow in the inset showing the dispersion of $q_2$. Pink dashed line in (c) is a visual guide showing the dispersion. (d) Energy dependence of QPI patterns near Fermi energy. (e) Direct comparison between FTs of d$I$/d$V$(**r**, V) maps at +16 mV (top half) and -2 mV (bottom half). (f, g) STM topograph and simultaneously taken d$I$/d$V$(**r**, -4 mV) map over a region encompassing a domain boundary denoted by a dashed white curve. Insets in (f,g) on each side of the boundary show a 2-fold symmetrized FTs of the corresponding region, demonstrating the rotation of the electronic signal. For visual purposes, all FTs have been two-fold symmetrized along the 1Q-CO wave vector (see Fig. S5 for unsymmetrized FTs). STM setup condition: (b) $V_{sample}$ = -4 mV, $I_{set}$ = 60 pA, $V_{exc}$ = 1 mV; (d) $V_{sample}$ = 18 mV, $I_{set}$ = 90 pA, $V_{exc}$ = 1 mV; $V_{sample}$ = 14 mV, $I_{set}$ = 100 pA, $V_{exc}$ = 1 mV; $V_{sample}$ = 12 mV, $I_{set}$ = 90 pA, $V_{exc}$ = 1 mV; $V_{sample}$ = 10 mV, $I_{set}$ = 70 pA, $V_{exc}$ = 1 mV; $V_{sample}$ = 5 mV, $I_{set}$ = 60 pA, $V_{exc}$ = 1 mV; $V_{sample}$ = -3 mV, $I_{set}$ = 60 pA, $V_{exc}$ = 1 mV; $V_{sample}$ = -6 mV, $I_{set}$ = 60 pA, $V_{exc}$ = 1 mV; $V_{sample}$ = -8 mV, $I_{set}$ = 80 pA, $V_{exc}$ = 1 mV; $V_{sample}$ = -10 mV, $I_{set}$ = 120 pA, $V_{exc}$ = 1 mV, T = 4.5 K; (e) $V_{sample}$ = 16 mV, $I_{set}$ = 200 pA, $V_{exc}$ = 1 mV; $V_{sample}$ = -2 mV, $I_{set}$ = 40 pA, $V_{exc}$ = 1 mV; (f,g) $V_{sample}$ = -4 mV, $I_{set}$ = 60 pA, $V_{exc}$ = 1 mV.

# References

1. Sachdev, S. Kagome- and triangular-lattice Heisenberg antiferromagnets: Ordering from quantum fluctuations and quantum-disordered ground states with unconfined bosonic spinons. *Phys. Rev. B* **45**, 12377–12396 (1992).

2. Mazin, I. I. *et al.* Theoretical prediction of a strongly correlated Dirac metal. *Nat. Commun.* **5**, 4261 (2014).

3. Guo, H.-M. & Franz, M. Topological insulator on the kagome lattice. *Phys. Rev. B* **80**, 113102 (2009).

4. Bilitewski, T. & Moessner, R. Disordered flat bands on the kagome lattice. *Phys. Rev. B* **98**, 235109 (2018).

5. Balents, L., Fisher, M. P. A. & Girvin, S. M. Fractionalization in an easy-axis Kagome antiferromagnet. *Phys. Rev. B* **65**, 224412 (2002).

6. Neupert, T., Santos, L., Chamon, C. & Mudry, C. Fractional Quantum Hall States at Zero Magnetic Field. *Phys. Rev. Lett.* **106**, 236804 (2011).

7. Plat, X., Alet, F., Capponi, S. & Totsuka, K. Magnetization plateaus of an easy-axis kagome antiferromagnet with extended interactions. *Phys. Rev. B* **92**, 174402 (2015).

8. Wen, J., Rüegg, A., Wang, C.-C. J. & Fiete, G. A. Interaction-driven topological insulators on the kagome and the decorated honeycomb lattices. *Phys. Rev. B* **82**, 075125 (2010).

9. Yu, S.-L. & Li, J.-X. Chiral superconducting phase and chiral spin-density-wave phase in a Hubbard model on the kagome lattice. *Phys. Rev. B* **85**, 144402 (2012).

10. Kiesel, M. L., Platt, C. & Thomale, R. Unconventional Fermi Surface Instabilities in the Kagome Hubbard Model. *Phys. Rev. Lett.* **110**, 126405 (2013).

11. Sun, K., Gu, Z., Katsura, H. & Das Sarma, S. Nearly Flatbands with Nontrivial Topology. *Phys. Rev. Lett.* **106**, 236803 (2011).

12. Tang, E., Mei, J.-W. & Wen, X.-G. High-Temperature Fractional Quantum Hall States. *Phys. Rev. Lett.* **106**, 236802 (2011).

13. O'Brien, A., Pollmann, F. & Fulde, P. Strongly correlated fermions on a kagome lattice. *Phys. Rev. B* **81**, 235115 (2010).

14. Rüegg, A. & Fiete, G. A. Fractionally charged topological point defects on the kagome lattice. *Phys. Rev. B* **83**, 165118 (2011).

15. Yan, S., Huse, D. A. & White, S. R. Spin-Liquid Ground State of the S = 1/2 Kagome Heisenberg Antiferromagnet. *Science* **332**, 1173–1176 (2011).

16. Isakov, S. V., Wessel, S., Melko, R. G., Sengupta, K. & Kim, Y. B. Hard-Core Bosons on the Kagome Lattice: Valence-Bond Solids and Their Quantum Melting. *Phys. Rev. Lett.* **97**, 147202 (2006).

17. Wang, W.-S., Li, Z.-Z., Xiang, Y.-Y. & Wang, Q.-H. Competing electronic orders on

kagome lattices at van Hove filling. *Phys. Rev. B* **87**, 115135 (2013).

18. Jiang, H.-C., Devereaux, T. & Kivelson, S. A. Holon Wigner Crystal in a Lightly Doped Kagome Quantum Spin Liquid. *Phys. Rev. Lett.* **119**, 067002 (2017).
19. Wang, Q. *et al.* Large intrinsic anomalous Hall effect in half-metallic ferromagnet $Co_3Sn_2S_2$ with magnetic Weyl fermions. *Nat. Commun.* **9**, 3681 (2018).
20. Morali, N. *et al.* Fermi-arc diversity on surface terminations of the magnetic Weyl semimetal $Co_3Sn_2S_2$. *Science* **365**, 1286–1291 (2019).
21. Yin, J.-X. *et al.* Negative flat band magnetism in a spin–orbit-coupled correlated kagome magnet. *Nat. Phys.* **15**, 443–448 (2019).
22. Jiao, L. *et al.* Signatures for half-metallicity and nontrivial surface states in the kagome lattice Weyl semimetal $Co_3Sn_2S_2$. *Phys. Rev. B* **99**, 245158 (2019).
23. Liu, E. *et al.* Giant anomalous Hall effect in a ferromagnetic kagome-lattice semimetal. *Nat. Phys.* **14**, 1125–1131 (2018).
24. Lin, Z. *et al.* Dirac fermions in antiferromagnetic FeSn kagome lattices with combined space inversion and time-reversal symmetry. *Phys. Rev. B* **102**, 155103 (2020).
25. Kang, M. *et al.* Dirac fermions and flat bands in the ideal kagome metal FeSn. *Nat. Mater.* **19**, 163–169 (2020).
26. Lin, Z. *et al.* Flatbands and Emergent Ferromagnetic Ordering in $Fe_3Sn_2$ Kagome Lattices. *Phys. Rev. Lett.* **121**, 096401 (2018).
27. Yin, J.-X. X. *et al.* Giant and anisotropic many-body spin–orbit tunability in a strongly correlated kagome magnet. *Nature* **562**, 91–95 (2018).
28. Ortiz, B. R. *et al.* New kagome prototype materials: discovery of $KV_3Sb_5$, $RbV_3Sb_5$, and $CsV_3Sb_5$. *Phys. Rev. Mater.* **3**, 094407 (2019).
29. Ortiz, B. R. *et al.* $CsV_3Sb_5$: A Z2 Topological Kagome Metal with a Superconducting Ground State. *Phys. Rev. Lett.* **125**, 247002 (2020).
30. Yang, S.-Y. *et al.* Giant, unconventional anomalous Hall effect in the metallic frustrated magnet candidate, $KV_3Sb_5$. *Sci. Adv.* **6**, eabb6003 (2020).
31. Ortiz, B. R. *et al.* Superconductivity in the $Z_2$ kagome metal $KV_3Sb_5$. *ArXiv*:2012.09097 (2020).
32. Wang, D. *et al.* Evidence for Majorana bound states in an iron-based superconductor. *Science* **362**, 333–335 (2018).
33. Zhang, P. *et al.* Observation of topological superconductivity on the surface of an iron-based superconductor. *Science* **360**, 182–186 (2018).
34. Wang, Y. *et al.* Proximity-induced spin-triplet superconductivity and edge supercurrent in the topological Kagome metal $K_{1-x}V_3Sb_5$. *ArXiv*:2012.05898 (2020).
35. Zhao, C. C. *et al.* Nodal superconductivity and superconducting dome in the topological

Kagome metal CsV$_3$Sb$_5$. *Arxiv*:2102.08356 (2021).

36. Jiang, Y.-X. *et al.* Discovery of topological charge order in kagome superconductor KV$_3$Sb$_5$. *ArXiv*:2012.15709 (2020).

37. Kostin, A. *et al.* Imaging orbital-selective quasiparticles in the Hund's metal state of FeSe. *Nat. Mater.* **17**, 869–874 (2018).

38. Fernandes, R. M. & Venderbos, J. W. F. Nematicity with a twist: Rotational symmetry breaking in a moiré superlattice. *Sci. Adv.* **6**, eaba8834 (2020).

39. Kerelsky, A. *et al.* Maximized electron interactions at the magic angle in twisted bilayer graphene. *Nature* **572**, 95–100 (2019).

40. Jiang, Y. *et al.* Charge order and broken rotational symmetry in magic-angle twisted bilayer graphene. *Nature* **573**, 91–95 (2019).

41. Xie, Y. *et al.* Spectroscopic signatures of many-body correlations in magic-angle twisted bilayer graphene. *Nature* **572**, 101–105 (2019).

42. Choi, Y. *et al.* Electronic correlations in twisted bilayer graphene near the magic angle. *Nat. Phys.* **15**, 1174–1180 (2019).

43. Fradkin, E., Kivelson, S. A. & Tranquada, J. M. Colloquium : Theory of intertwined orders in high temperature superconductors. *Rev. Mod. Phys.* **87**, 457–482 (2015).

**Methods**

Single crystals of CsV$_3$Sb$_5$ were grown and characterized as described in more detail in Ref. [29]. We cold-cleaved and studied three different CsV$_3$Sb$_5$ crystals, all of which exhibited qualitatively the same phenomena described in the main text (Fig. S8). STM data was acquired using a customized Unisoku USM1300 microscope at varying temperatures denoted in figure captions. Spectroscopic measurements were made using a standard lock-in technique with 915 Hz frequency and bias excitation as also detailed in figure captions. STM tips used were home-made chemically-etched tungsten tips, annealed in UHV to bright orange color prior to STM experiments.

.